# Evaluation of Classical Features and Classifiers in Brain-Computer Interface Tasks

Ehsan Arbabi [1,2] and Mohammad Bagher Shamsollahi [1]

[1] Department of Electrical Engineering, Sharif University of Technology, Iran
[2] School of Electrical and Computer Engineering, College of Engineering, University of Tehran, Iran

earbabi@ut.ac.ir, mbshams@sharif.edu

**Abstract: Brain-Computer Interface (BCI) uses brain signals in order to provide a new method for communication between human and outside world. Feature extraction, selection and classification are among the main matters of concerns in signal processing stage of BCI. In this article, we present our findings about the most effective features and classifiers in some brain tasks. Six different groups of classical features and twelve classifiers have been examined in nine datasets of brain signal. The results indicate that energy of brain signals in α and β frequency bands, together with some statistical parameters are more effective, comparing to the other types of extracted features. In addition, Bayesian classifier with Gaussian distribution assumption and also Support Vector Machine (SVM) show to classify different BCI datasets more accurately than the other classifiers. We believe that the results can give an insight about a strategy for blind classification of brain signals in brain-computer interface.**

*Keywords: Brain-Computer Interface; Classification; Effective Feature; Electroencephalography (EEG); Feature Selection.*

## 1  Introduction

Brain-computer interface (BCI) provides a new way of communication between human and the outside world, by using brain signals (such as EEG, MEG, ECoG). Providing this kind of communication can have many applications, especially for aiding people with major physical disabilities. In fact, by processing brain signals, the brain activities can be interpreted, and based on the interpretation a suitable action is performed artificially [1-5].

In order to obtain desired results in BCI-related applications, we need to have a correct interpretation of the brain signals. Therefore, special attention should be paid to the signal processing stage. The signal processing stage of the brain signal can be usually divided into three sub-stages: 1- Feature extraction from brain signals, 2- Feature reduction/selection, and 3- Classification of the recorded brain signals based on the selected features. First of all, we need to extract the features which seem to be potentially efficient for our task. Then, we may reduce size of the extracted feature by selecting the most effective ones and/or combing them with each other. At last, a classifier can be applied to use these effective features for labeling the recorded signals based on their related brain activity [6].

There are too many works and research in the area of BCI, which are presented in different scientific articles. Many of these works are mainly focusing on special tasks of BCI. In fact, they usually present their strategy for processing brain signals and demonstrate the power of their suggested methods by testing them in special BCI tasks [2, 3, 7, 8]. Therefore, several methods of signal processing which are mainly different in the mentioned three sub stages have been presented. Although, all these works are providing valuable information about processing brain signals in specific tasks, they do not provide a straightforward strategy for general tasks. In other words, they cannot say which features and which classifiers seem to be potentially the most suitable ones for completely new BCI tasks. Of course, finding an absolute answer to the concern mentioned above may be very hard, if not impossible. However, we can still have a general straightforward strategy when facing new BCI projects if at least we know which features and classifiers seem to be potentially more effective for BCI tasks (among the typical features and classifiers).

The aim of this study, which is originally done in 2005 in Department of Electrical Engineering at Sharif University of Technology, is to find a general answer to the mentioned concern, rather than just suggesting another new method for a specific BCI task or even comparing the solutions in some limited cases [9, 10]. Therefore, instead of evaluating a new method of pattern recognition in BCI, we try to find relation between the most effective features and classifiers in nine BCI tasks. In order to obtain a general answer, we process each BCI dataset in the same manner, and avoid adapting our processing strategy to the characteristics of each BCI task separately. We start by extracting different typical features from brain signals. Then, due to the very large size of extracted features, we reduce the size of feature vectors without changing/combining them. In the next step, we train and test twelve different classifiers with these extracted features. Finally, by exploiting some searching methods, we find the selections of effective features which work well with the applied classifiers.

## 2   Material and methods

### 2.1 Data Acquisition

We worked on datasets which are available online for research. The nine datasets are EEG signals provided by IDIAP Research Institute (Silvia Chiappa, José del R. Millán). These datasets contain data from three normal subjects during a few non-feedback sessions. The subjects sat in a normal chair, relaxed arms resting on their legs. There were three tasks:
1. Imagination of repetitive self-paced left hand movements,
2. Imagination of repetitive self-paced right hand movements,
3. Generation of words beginning with the same random letter.

Sampling rate was 512 Hz and the data were split to 1 second segments, where each segment was overlapped with the next and the previous segments by 0.5 second. All the data of a given subject were acquired on the same day. In each session, the subject performed a given task for about 15 seconds and then switched randomly to another task at the operator's request [11]. Totally 658 seconds, 650 seconds and 643 seconds of trials were considered for the 1st, 2nd and 3rd subject, respectively. By having three BCI tasks for three subjects we could have nine different dual BCI tasks (Table 1). The EEG channels used for these data recording are: Fp1, AF3, F7, F3, FC1, FC5, T7, C3, CP1, CP5, P7, P3, Pz, PO3, O1, Oz, O2, PO4, P4, P8, CP6, CP2, C4, T8, FC6, FC2, F4, F8, AF4, Fp2, Fz and Cz.

**Table 1** BCI datasets: number 1 to number 9

| No. of Dataset | Subject | Task 1 | Task 2 | Task 3 |
|---|---|---|---|---|
| 1 | One | X | X | |
| 2 | one | X | | X |
| 3 | one | | X | X |
| 4 | two | X | X | |
| 5 | two | X | | X |
| 6 | two | | X | X |
| 7 | three | X | X | |
| 8 | three | X | | X |
| 9 | three | | X | X |

### 2.2 Pre-Processing

The signals were filtered between 0.5 Hz and 45 Hz for removing the possible additive noises such as offsets and 50/60 Hz power line noise. Also, in order to have faster processing, the recorded signals were down sampled from from 512 Hz to 128 Hz.

### 2.3 Feature Extraction

After pre-processing, for every trial (considering all channels for EEG signals), different types of feature were extracted from the brain signals. The extracted groups of features are listed below.

*a)      Group 1- Statistical features*

Joint Central Moment of all channels (order 1 to 5); Joint Central Cumulant between 2, 3 and 4 channels (order 1 to 5); Correlation between all channels; Form Factor for each channel separately; Statistical Variance for each channel separately [12, 13].

*b)      Group 2- Entropy-based features*

Approximated Neural Complexity Measure between all channels; Lample-Ziv Complexity Measure for each channel separately; Approximation Entropy for each channel separately; Shannon, Renyi and Tsallis Entropies (for α and q = -5, -2, -1, 0.5, 1.5, 2, 3, 5) [14-17].

*c)*   *Group 3- Estimated parameters of Autoregressive model (AR) of brain signals*

AR model of order 4; AR model of order 8; AR model of order 16; AR model of order 32 [18].

*d)*   *Group 4- Signal Energy in different frequency ranges*

The energy of brain signals in different frequency ranges constructs our fourth group of features. The bandwidth range of EEG signal is from DC to 100 Hz. However, the major power is distributed in the lower half of this range. Usually, four major frequency bands of brain signals are considered as δ (0.5 Hz to 4 Hz), θ (4 Hz to 8 Hz), α (8 Hz to 13 Hz) and β (13 Hz to 22 Hz [18].

*e)*   *Group 5- DCT and DST coefficients*

Coefficients of Discrete Cosine Transform; Coefficients of Discrete Sinusoid Transform [19].

*f)*   *Group 6- Wavelet Transform coefficients*

Coefficients of Haar Wavelet Transform; Coefficients of Daubechies 2 Wavelet Transform; Coefficients of Daubechies 3 Wavelet Transform; Coefficients of Daubechies 4 Wavelet Transform; Coefficients of Daubechies 5 Wavelet Transform [20].

After extracting the features, they were normalized in order to prevent numerical computational errors.

## 2.4 Omitting Irrelevant Features

Due to large number of extracted features (thousands of features for each trial), we had to reduce their size. Since we wanted to find effective features, we only applied feature selection methods and avoided using the feature reduction methods which may change the original essence of features (such as PCA or ICA). We performed two different feature selection methods for each group of features (group 1 to group 6), separately.

### 2.4.1   Classifier-Independent Method

In this method, three evaluation measures were calculated for all the features: 1- Mahalanobis distance, 2- Bhattacharyya distance, and 3- the measure based on scattering matrices [21]. After normalizing them, their sum was calculated (for each feature separately). Based on the sum values the features were ranked. In order to reduce the effect of considering similar features with different names, we also checked the correlation between them. In other words, if a feature is correlated with a higher ranked feature, its ranking would become lowered based on the correlation coefficient.

As it can be seen, these measures are independent from classification method. They only evaluate how much the extracted features can separate the related classes. We applied this method of feature reduction, because it is faster than classifier-dependent methods (explained below). But, since the results of such classifier-independent method could not be satisfactory, we only used it for having a pre-ranking of the features. After that, we performed a classifier-dependant feature reduction method on the 200 highest pre-ranked features.

### 2.4.2   Classifier-Dependent Method

We used twelve classifiers for evaluating different groups of features in the explained BCI tasks. List of these classifiers and their brief description can be seen in Table 2. Test and train method was applied for estimating the accuracy of classification (correctness of the classification). In this method, the training was performed by using 2/3 of samples, and then we evaluated the performance of the classifiers by testing the trained classifiers on the other 1/3 of samples.

To find the most effective subset of features inside each group of features, the evaluation measure was defined as the calculated accuracy of the classifier. A search was performed on the top 200 pre-ranked features (pre-ranked by classifier-independent method) to find top 20 effective features among them (by using classification measure), for each group of features separately.

In order to have a complete search, it is necessary to evaluate all possible subsets of features by the defined evaluation measure. Therefore, in order to find 20 effective features from a total number of 200 features, 20-combination of 200 or C(200, 20) number of searches is needed. It can be realized that using optimal searching method can be extremely slow and time consuming (C(200, 20) = $1.6 \times 10^{27}$). Therefore, instead of performing a complete search, suboptimal searching methods like Genetic searching algorithm and Floating Sequential Forward Selection (FSFS) was applied. Genetic algorithm generates new generations of features by using the best features in the mother generations. The mother generations are usually chosen randomly [26]. FSFS starts from an empty subset. Then, it generates new subsets in each iteration, by adding a feature selected by the evaluation measure [21].

**Table 2** The applied classifiers and their brief description

| Classifier | Brief Description |
| --- | --- |
| Bayes | Bayesian classifier with Gaussian probability density assumption for extracted features [18] |
| SVM | Linear Support Vector Machine [22, 23] |
| Percep | One layer Perceptron (Artificial Neural Networks) [21] |
| MLP2TG | Two layer Perceptron with Tan-Sigmoid function in nodes (Artificial Neural Networks) [21] |
| MLP2PN | Two layer Perceptron with Purelin function in nodes (Artificial Neural Networks) [21] |
| MLP3TG | Three layer Perceptron with Tan-Sigmoid function in nodes (Artificial Neural Networks) [21] |
| MLP3PN | Three layer Perceptron with Purelin function in nodes (Artificial Neural Networks) [21] |
| RBF | Radius Basis Function classifier [24] |
| ANFIS1 | Type three Adaptive-Network-based Fuzzy Inference Systems with Gaussian membership function [25] |
| ANFIS2 | Type three Adaptive-Network-based Fuzzy Inference Systems with power-sigmoid membership function [25] |
| ANFIS3 | Type three Adaptive-Network-based Fuzzy Inference Systems with trapezoid membership function [25] |
| N-F-CM | Neuro-Fuzzy classifier using Fuzzy C-mean algorithm and two layer Perceptron with Tan-Sigmoid function in nodes [21] |

After finding the most effective features in each group of features, another search was performed to find top 25 effective features among the bests of all groups (for each classifier separately). Because of very slow performance of ANFIS classifier, we found the 5th most effective features for this classifier, instead of 20 or 25.

## 3 Results

In each dataset, for every classifier, the percentage of correct classification by using the top selected features of each group was obtained. The effect of different groups of features on Bayesian and SVM classifiers in each dataset is presented in Fig. 1 and Fig. 2, respectively. Similar figures are also available for the other classifiers; however, we only illustrated them for Bayesian and SVM classifiers because of their better classification.

In order to have more general view of the relation between effective features and classifies, with less dependency on the datasets, we calculated the average and standard deviation of each value (classification accuracy) in all nine datasets. In other words, for each set of features (6 groups + 1 best of all) the average percentage of correct classification by using each classifier, and its standard deviation were obtained. The results are presented in Table 3.

**Table 3** Average and standard deviation of classification accuracy in all datasets, by using different groups of features and different classifiers (average / standard deviation). The dark gray cells represent the groups of features which give the best accuracy in the related classifier (1st place). The light gray cells represent the groups of features which give next to the best accuracy in the related classifier (2nd place). "Best of all groups" is related to effective features selected from all groups of features together

| Classifier | Feature Group | | | | | | |
| --- | --- | --- | --- | --- | --- | --- | --- |
| | Group 1 (Statistic) | Group 2 (Entropy) | Group 3 (AR) | Group 4 (Energy) | Group 5 (DCT, DST) | Group 6 (Wavelet) | Best of all groups |
| BAYES | 82.2 / 4.4 | 77.7 / 6.3 | 77.3 / 6.0 | 78.0 / 7.5 | 79.4 / 6.0 | 79.5 / 5.6 | 86.1 / 4.4 |
| SVM | 82.1 / 6.4 | 76.4 / 5.3 | 77.5 / 5.9 | 80.2 / 7.6 | 71.7 / 3.7 | 70.6 / 2.8 | 85.4 / 3.9 |
| RBF | 68.7 / 7.5 | 65.3 / 6.2 | 64.9 / 4.9 | 71.7 / 9.5 | 64.6 / 4.8 | 64.6 / 4.2 | 71.9 / 8.3 |
| PERCEP | 70.9 / 5.1 | 66.8 / 4.7 | 67.3 / 6.1 | 72.8 / 10.7 | 61.1 / 4.4 | 63.3 / 2.7 | 75.1 / 4.7 |
| MLP2tg | 67.4 / 6.1 | 62.8 / 7.2 | 62.2 / 6.0 | 68.5 / 8.0 | 63.3 / 5.1 | 61.6 / 4.1 | 69.4 / 6.1 |
| MLP2pn | 72.9 / 5.6 | 68.6 / 5.8 | 69.7 / 7.1 | 75.0 / 8.5 | 66.5 / 2.5 | 66.1 / 2.2 | 79.4 / 4.6 |
| MLP3tg | 66.3 / 7.1 | 61.9 / 7.5 | 62.9 / 6.6 | 69.7 / 7.6 | 62.0 / 5.2 | 61.5 / 4.4 | 70.0 / 8.0 |
| MLP3pn | 71.5 / 5.5 | 68.3 / 5.9 | 68.8 / 5.3 | 74.2 / 8.2 | 64.6 / 2.7 | 65.4 / 2.7 | 76.5 / 5.3 |
| N-F-CM | 63.9 / 5.5 | 61.3 / 6.0 | 59.0 / 3.4 | 68.8 / 8.5 | 59.0 / 4.3 | 59.8 / 2.7 | 65.4 / 5.8 |
| ANFIS1 | 67.4 / 5.7 | 63.6 / 5.2 | 61.6 / 3.8 | 72.2 / 8.1 | 65.0 / 5.4 | 62.3 / 3.3 | 71.9 / 7.0 |
| ANFIS2 | 67.3 / 6.5 | 63.0 / 5.8 | 63.4 / 4.6 | 71.8 / 7.8 | 64.3 / 6.0 | 62.5 / 3.7 | 71.3 / 7.0 |
| ANFIS3 | 67.2 / 4.1 | 62.6 / 6.4 | 63.6 / 5.3 | 73.3 / 7.9 | 64.7 / 6.9 | 62.7 / 3.6 | 72.8 / 6.4 |

As it was explained before, each group of features consists of several different features. For example, the first group of features included different statistical parameters: moments, cumulant, correlation, form factor and variance. As an instance, Fig. 3 shows the distribution of different types of statistical features inside the related group (group 1), for some of the exploited classifiers (for the 2nd dataset).

It was important to know which features, inside each group of features, are providing the best classification result for each classifier. Since features from Group 1 (Statistic) and Group 4 (Energy) provide the best classification results, we found the more effective individual features inside these groups (comparing to the other individual features):
- Signal energies between 8 Hz to 15 Hz and also around 20 Hz (α and β bands) were among the most effective features in Group 4 (Energy).
- Signal energies in more than 30 Hz was usually the least effective features in Group 4 (Energy).
- Correlation between recording channels was the most effective feature in Group 1 (Statistic). After correlation, form factor could also sometimes be considered as effective feature (to less extend comparing to correlation).

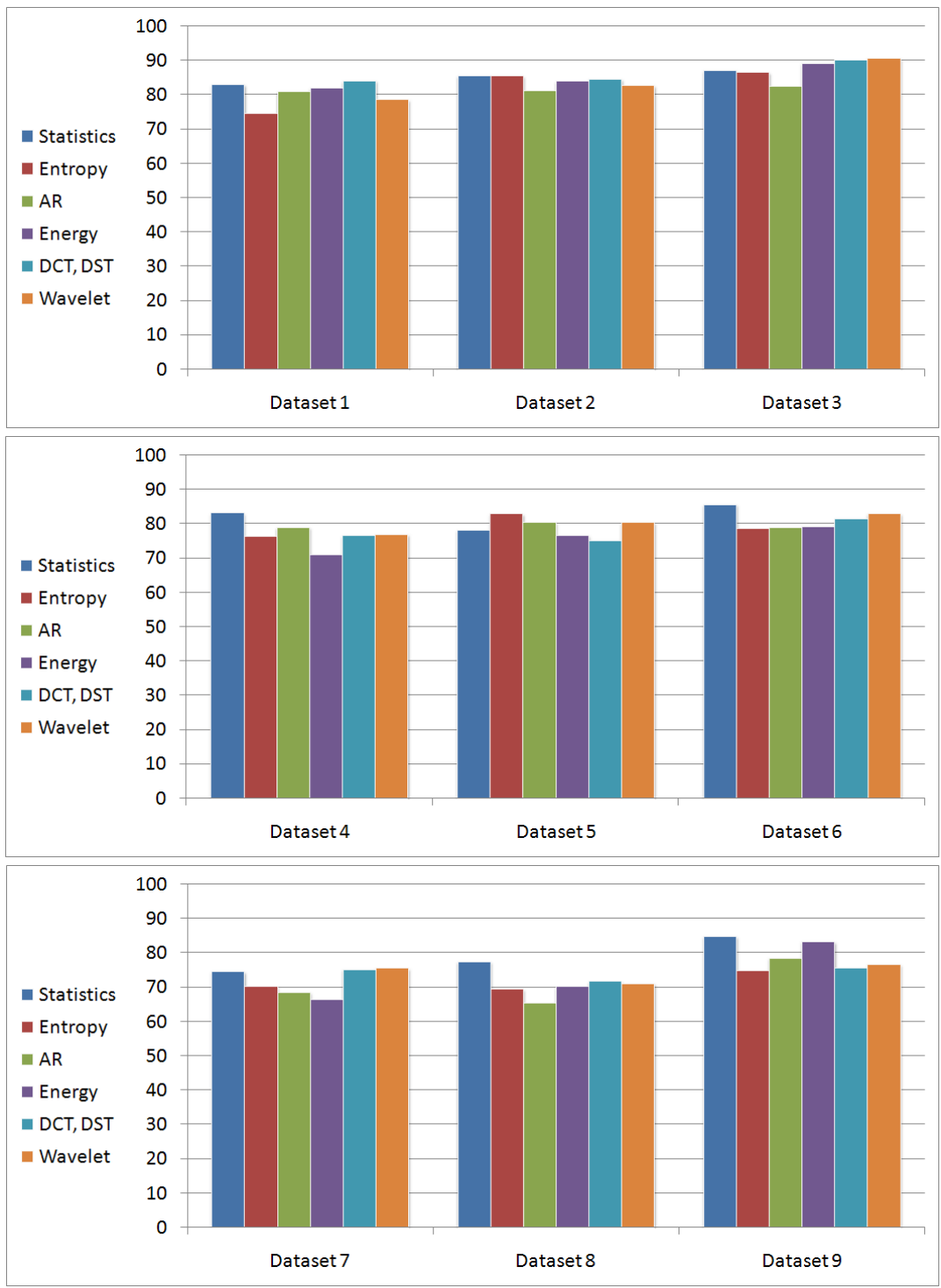

Fig. 1 The effect of different groups of features on Bayesian classifier (percentage of correct classification) in each dataset

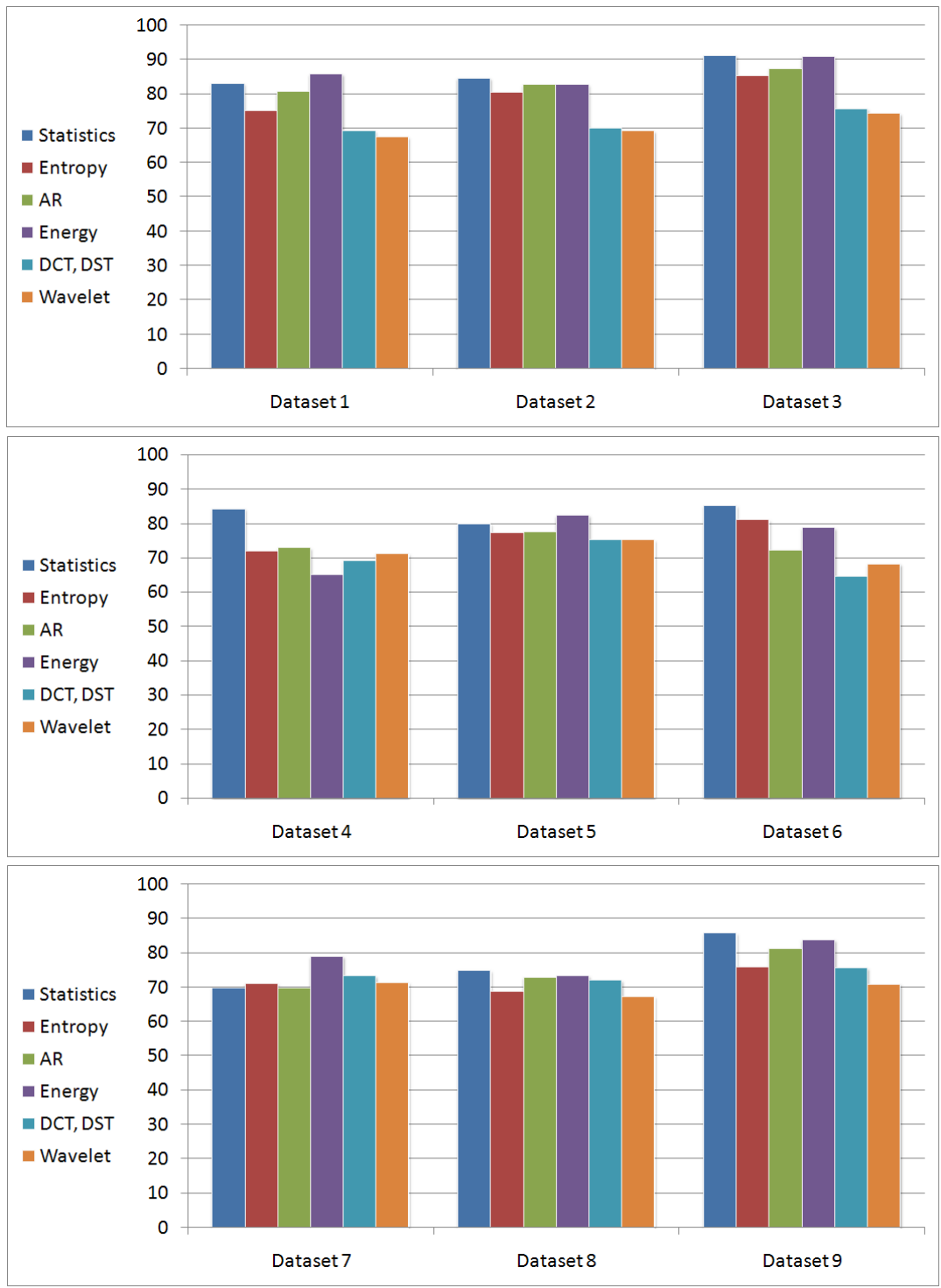

**Fig. 2** The effect of different groups of features on SVM classifier (percentage of correct classification) in each dataset

## 4  Discussion

As it could be understood from the results, in average, the traditional features such as signal energy in different frequency bands and statistical factors are providing the best classification results, for the exploited datasets. Of course, it does not mean that the other groups of features (such as DCT/DST or Wavelet-based features) are always less effective than statistical and energy. In other words, for classifying brain data, using statistical and energy features can more probably result in a better classification, comparing to other features. Excepting Bayesian classifier, the best and next to the best groups of features for all classifiers are energy and statistical features. The classification accuracy by using the best group of features (dark gray cells in Table 3) is in average 2.8% (std = 1.6%) more than the classification accuracy by using next to the best group of features (light gray cells in Table 3).

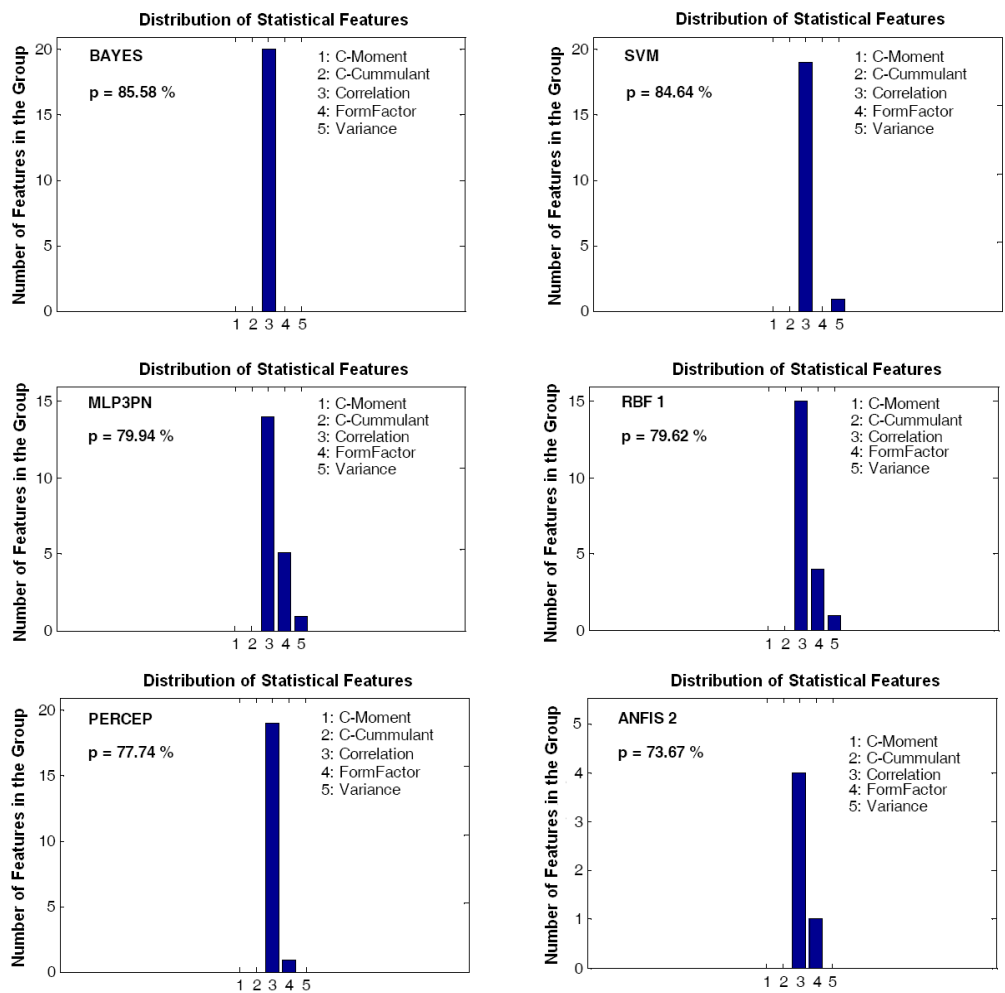

**Fig. 3** Distribution of different types of statistical features inside statistical group (for the 2$^{nd}$ dataset, as an example)

Inside energy group of features, signal energies in α band and to less extend in β band are in average more effective than signal energy in the other bands. Especially, signal energy in more than 30 Hz was usually the least effective in classification. Among statistical group of features, correlation between recorded channels could provide more useful information for classification. It can suggest that correlation between brain activities in different parts of brain can be a good clue for interpreting the mental task (comparing to the other statistical parameters). In case we only have single channel, using form factor seems to be more effective for classification.

Among all the classifiers, Bayesian classifier with Gaussian distribution assumption and SVM classifier were in average performing better than the other classifiers. Better performance of the applied Bayesian classifier can suggest that extracted effective features may be distributed more similar to Gaussian function. In addition to accuracy, it should be noticed that the applied Bayesian classifier is based on simple conditional probability equations. Therefore, it is much faster than many other classifiers, such Artificial Neural Networks and ANFIS classifiers.

While energy features are among the most effective features for most of the classifiers, DCT/DST and Wavelet-based features are in average providing better classification results for Bayesian classifier (in addition to statistical features). This may suggest that DCT/DST and Wavelet-based features of brain signals have more suitable probability distribution for Gaussian-based Bayesian classifier, comparing to energy features.

When we select 25 effective features from all types features (i.e. considering all groups of features together), it is expected that the classification accuracy increases. When effective features from all types are considered (most right column in Table 3), in average, the classification accuracy is only 0.7% higher than when the best group of effective features is considered (dark gray cells in Table 3) (with std = 2.5%). Of course, for some classifiers the accuracy decreases, which can be due to using a sub-optimal searching method (rather than slow optimal searching method). If we neglect such decreases, the classification accuracy by the best of all features (all types) would be in average 1.4% (instead of 0.7%) more than the classification accuracy of the best group of features (with std = 1.6%). This shows that when we consider only the best group of features (e.g. only statistical features), the classification results are almost as good as when we consider all groups of features together.

As it was discussed before, during evaluating different groups of features separately, statistical and energy features provided better classification results, comparing to the other groups. However, when we selected the effective features from all types of feature, we observed that statistical and energy features are not necessarily acting more significantly than the others. This indicates that although considering all types of feature (together) in a typical BCI task may provide a good classification result, it may not give a correct idea about the best type of feature for that task.

## 5 Conclusion

In this paper, we compared the effect of different types of feature and classifiers for classifying brain signals in nine different BCI tasks. In order to be able to generalize our results, we avoided adapting our processing strategy to each BCI dataset separately. Instead, we treated all datasets equally. Six groups of features were extracted from brains signals (EEG) for all nine datasets. Then, twelve different classifiers were used for finding the most effective features among these groups. The results show that statistical features and signal energy in different frequency bands are among the most effective feature in all tasks. Especially, signal energy in $\alpha$ band and correlation between different recording channels result in comparatively higher classification results. Among the applied classifiers, Bayesian classifier with Gaussian distribution assumption and also SVM classifier performed the best.

The results were obtained based on different tasks in different individuals. Therefore, they can give a general view about classification tasks in BCI problems. In fact, the results suggest that when facing a new task in BCI, it is strongly recommended not to miss extracting statistical features (esp. correlation between recorded channels and form factor) and energy features (especially in $\alpha$ and $\beta$ bands) from the signals. Using Bayesian classifier is recommended due to its simple implementation, fast performance and comparatively better classification. It should be also considered that DCT/DST and Wavelet coefficients can also improve the accuracy of classification, especially when using Bayesian classifiers with Gaussian distribution assumption.

Evaluating large number of features and classifiers in different BCI datasets is a very time-consuming task. However, the results of this work can still be extended if we continue it by evaluating even more groups of features in more BCI tasks.